\documentclass[12pt,a4paper]{scrartcl}
\usepackage[utf8]{inputenc}   
\usepackage[T1]{fontenc}  
\usepackage[english]{babel} 
\usepackage[usenames,dvipsnames]{xcolor}      
\usepackage[cm]{aeguill} 
\usepackage{changes} 
\usepackage{alltt,amsbsy,amsmath,amssymb,amsthm,amsopn,amstext,amscd,amsfonts,amsxtra,array,bbding,bbm,cancel,dsfont,epic,latexsym,nicefrac,pifont} 
\usepackage{enumerate,multicol,multirow,verbatim}
\usepackage{graphicx,epsfig,xcolor}
\usepackage{grffile} 

\usepackage{subfigure}
\usepackage{xfrac}  
\usepackage{tikz}         
\usepackage{stmaryrd} 
\usepackage{xpatch} 
\usepackage[linesnumbered,ruled,noline,commentsnumbered,longend]{algorithm2e}
\usepackage{hyperref}  
\hypersetup{colorlinks=true,pagebackref=true,urlcolor=orange,citecolor=brown}

\newtheorem{remark}{\textit{Remark}}[section]

\newtheorem{defi}{Definition}[section]
\newtheorem{Ex}{Example}[section]

\title{ToMATo: an efficient and robust clustering algorithm for high dimensional datasets. An illustration with spike sorting.}
\author{Louise Martineau, Christophe Pouzat, Ségolen Geffray}
\begin{document}

\maketitle

\section*{Abstract}
Clustering algorithms became an essential part of the neurophysiological data analysis toolbox in the last twenty five years. Many problems, from the definition of cell types/groups based on morphological, molecular and physiological data to the identification of sub-networks in fMRI data, are now routinely tackled with clustering analysis. Since the datasets to which this type of analysis is applied tend to be defined in larger and larger dimensional spaces, there is a need for efficient and robust clustering methods in high dimension. There is also a need for methods that assume as little as possible about the clusters shape and size. We report here our experience with the ToMATo (Topological Mode Analysis Tool) algorithm. It is based on a definitely deep mathematical theory (algebraic topology), but its Python based open-source implementation is easily accessible to practitioners. We applied ToMATo to a problem we know well, spike sorting. Its capability to work in the “native” space of the data (no dimension reduction is required) is remarkable, as well as its robustness with respect to outliers (superposed spikes).

\section{Introduction} 
Clustering method \cite[Chap. 14]{hastie.tibshirani.friedman:09} applications to neurobiological data have undergone a spectacular growth in recent years. To cite just a few examples: spike sorting \cite{pouzat:16,kiraly.balazs:22}; grouping voxels with similar activities from fMRI data \cite{aljobouri.ea:18}; identifying neural activity patterns from extracellular recordings \cite{berry.tkacik:20}. \\

Recording and decoding the activity of multiple neurons is a major subject in contemporary neuroscience. Extracellular recordings with multi-electrode arrays is one of the basic tools used to that end. The raw data produced by these recordings are almost systematically a mixture of activities from several neurons. In order to find the number of neurons which contributed to the recording and identify which neuron generated each of the visible spikes, a pre-processing step called spike sorting is required.\\

Spike sorting is nowadays a semi-automatic process which involves many steps. Indeed,
following some initial steps (data normalization, spike detection, event construction), spike sorting boils down to a clustering problem in high dimension.
It is therefore accompanied most of the time by a dimension reduction step. This
dimension reduction step is sensitive to the presence of superpositions, that is, the superposition of the activity of two or more neurons that fired nearly simultaneously. These superpositions, akin to outliers, lead to poor clustering results. Neuroscientists are then usually led to perform an extra pre-processing step to remove these superpositions. This step is not completely automated and does a much better job when supervised by an expert in spike sorting. In addition, many clustering methods exhibit a serious performance reduction with increasing dimension and the current trend is clearly towards data defined in larger and larger dimensional spaces (in spike sorting, but also for instance by considering longer time series).  Since clustering methods are typically used when little is \textbf{a priori} known about the data, it makes sense to use methods making very few hypothesis about the data, that is nonparametric methods. With these two requirements in mind we explored the performances of the ToMATo clustering method. An experienced reader might rightfully be surprised hearing about a method that is both nonparametric and able to cope with large dimensions. That is why we wrote that the data are \textbf{defined} in a space of large dimension, making a distinction between the vector length (dimension) defining each data point and the intrinsic dimension in which the data are living. If the data have a large intrinsic dimension, ToMATo may fail, but if this intrinsic dimension is small (of the order of 10), it performs very well (in our experience), even if the native space of the data has large dimension (in our example 180).\\

 The use of the ToMATo clustering algorithm helps to simplify and streamline this part of the spike sorting procedure.\\

ToMATo (Topological Mode Analysis Tool) is a clustering method using persistent homology, developed in 2013 by Chazal, Guibas, Oudot and Skraba \cite{Tomato}. It seems little known in neuroscience but we show in this article how it can be extremely helpful and effective using as an example an application to spike sorting. Indeed it enables to reduce the number of steps typically involved in this procedure while providing valuable results. Also, four main reasons to use the ToMATo algorithm for spike sorting are the following:  
\begin{enumerate}
\item ToMATo works without dimension reduction as a prior step.
\item ToMATo is robust to superpositions.
\item ToMATo provides a very easy way of choosing the right number of clusters, solving thereby a significant problem in clustering.
\item ToMATo runs fast and requires tuning only a few parameters. 
\end{enumerate}

Very importantly, the ToMATo algorithm is implemented and well-documented in Gudhi \footnote{\url{https://gudhi.inria.fr/python/latest/clustering.html.}}, a generic open source C++ library for topological data analysis, with a Python interface. It is thus very easy to use even for non specialists of persistent homology.\\

In section~\ref{tomato_section}, we present the ToMAto algorithm and explain how to use it in practice. This section is very detailed since we are convinced that the user of a method should have a clear understanding of how it is designed and how it works (ToMATo is not a black-box!). In section \ref{app}, we show applications of this algorithm to spike sorting on simulated and real data, demonstrating the spectacular performances of this approach. Appendix \ref{annexe} provides a high level description of a complete spike sorting procedure, in order to help the reader to see where ToMATo brings key improvements. 

\section{The ToMATo algorithm} \label{tomato_section}
\label{sec:tomato}
\subsection{Introduction} \label{intro_part}

The ToMATo (Topological Mode Analysis Tool) algorithm is a mode-seeking algorithm. The general idea of mode-seeking algorithms for clustering consists in seeking peaks in the observation density $f$, and in assigning observations falling under the same peak to the same cluster. Indeed, if points are sampled under $f$, there should be a cluster of points corresponding to each peak of $f$. This type of algorithms is supposed to be able to find clusters of any shape, as opposed to algorithms such as kmeans which work well only on convex clusters. A major problem classically arises in mode-seeking: it can be very sensitive to small perturbations of $f$. It turns out that in practice we have only access to an approximation $\hat{f}$ of the true density $f$ and that the peaks of $\hat{f}$ do not in general coincide with the ones of $f$.\\

Several strategies can be considered to address this issue. The innovative approach of ToMATo resides in the use of persistent homology theory, and thanks to persistent homology a notion of peak prominence is introduced, such that prominent peaks of $\hat{f}$ correspond to prominent peaks of $f$. Clusters found by mode-seeking are merged together so that the final clusters correspond only to prominent peaks of the true density $f$, and not to some spurious, noise induced, peaks of $\hat{f}$. \\

More precisely, ToMATo is based on a graph mode-seeking algorithm, introduced in \cite{Koontz}. It only uses a density estimate at the data points, and performs mode-seeking on an auxiliary structure, a neighbourhood graph, instead of performing mode-seeking directly on a density estimate. This graph-based mode-seeking method allows for making effective computations. It still suffers from the same drawback as any mode-seeking method: it is likely to find too many clusters by taking into account noise induced peaks.
To recover some stability, ToMATo therefore combines this computationally effective graph-based mode-seeking, with a merging step based on peak prominence.

\subsection{Ideas behind ToMATo in the continuous setting} \label{continuous}

To give insight and understand the ideas behind ToMATo, we first give an overview of what happens in the continuous setting, that is, in the theoretical case where we work directly on (smooth) functions. We start with some reminders on mode-seeking, and persistent homology. In the following we denote a generic function by $g$ and we reserve the notation $f$ for functions that are densities.
 
\subsubsection{Gradient ascent for mode-seeking}    

Let us recall the principle of gradient ascent mode-seeking. We highlight the fact that in practice in the discrete case, we won't need to estimate density gradients. We talk about gradient ascent here in the continuous setting, to mathematically define the intuitive notions of peaks and their ascending regions.\\

\begin{figure}    
\begin{center} 
\includegraphics[scale=0.25]{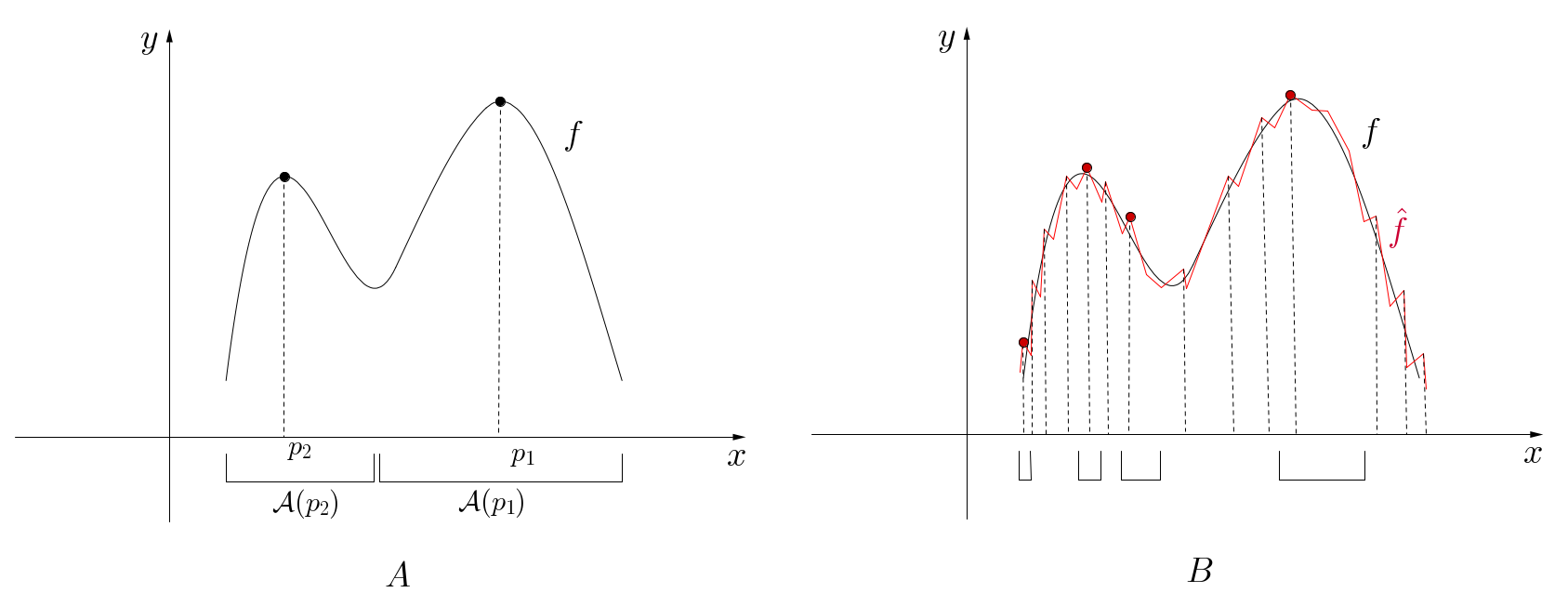}
\end{center} 
\caption{ A: A density function $f$ with two peaks $p_1$ and $p_2$. Their respective ascending regions are denoted by $\mathcal{A}(p_1)$ and $\mathcal{A}(p_2)$. B: An approximation $\hat{f}$ of $f$ based on a sample is represented in red. This approximation has many peaks, only a few are highlighted together with their ascending regions.} 
\label{Tomato1}
\end{figure}

Intuitively a cluster should be a group of data points being under the same peak of the (unknown) density $f$. In practice, a peak can be defined as a local maximum of $\hat{f}$, which can be identified by gradient ascent. We define the ascending region of a peak, as the set of data points that converge to this peak when used as starting points of the gradient ascent procedure. The clusters will then be defined as the ascending regions of $\hat{f}$. Ascending regions is an unstable quantity, as illustrated in Figure \ref{Tomato1}. 
 
\subsubsection{Persistent homology  of functions} 
\label{sec:persistent-homology} 
Persistent homology of functions is a method rooted in both Morse theory and topological data analysis. At the core of persistent homology  of a generic function $g$, is the evolution of the connected components \footnote{We can also be interested in the evolution of $1-$dimensional holes (the void inside a circle or a triangle), $2-$dimensional holes (the void inside a sphere or a tetrahedron) and so on. For clustering, we only need to look at connected components, which can be seen as $0-$dimensional holes.} of the superlevel sets of $g$, the definition of which is recalled below. 
\begin{defi} Superlevel set \\
Let $g$ be a function from $\mathbb{R}^d$ to $\mathbb{R}$ for $d \geq 1$. Let $\alpha \in \mathbb{R}$.
The superlevel set of $g$ of parameter $\alpha$ is:
\[ \lbrace x \, \vert \,  g(x) \geq \alpha \rbrace .\]
\end{defi}
 
\begin{figure}[h] 
\begin{center} 
\includegraphics[scale=0.25]{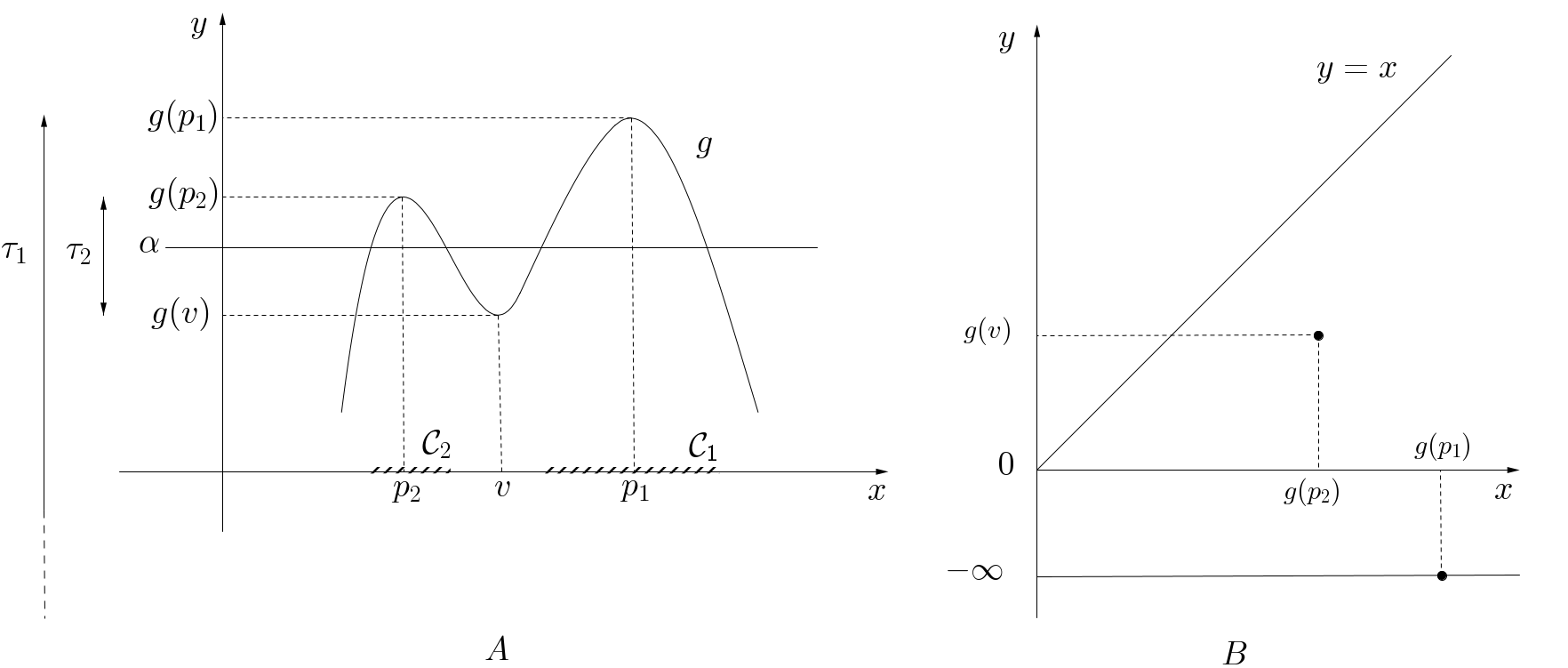}
\end{center}  
\caption{Persistent homology  of a function.} 
\label{Tomato2}
\end{figure}

We keep track of the connected components of $\lbrace x \, \vert \, g(x) \geq \alpha\rbrace$, as a parameter $\alpha$ decreases from $+ \infty$ to $- \infty$. We can imagine that $g$ represents an altitude and that $\alpha$ represents the sea level, then the connected components correspond to the surface (cross-sections) of the islands that appear as the sea level decreases. The nested sequence of sets $\left( \lbrace  x \, \vert \,g(x) \geq \alpha\rbrace \right)_{+ \infty \geq \alpha \geq -\infty}$ is called a \textbf{filtration}. Let us take the function in Figure \ref{Tomato2} A as an example. When $\alpha > g(p_1)$, $\lbrace x \, \vert \, g(x) \geq \alpha\rbrace$ is empty. Then, a connected component $\mathcal{C}_1$ appears at $\alpha=g(p_1)$, the global maximum of $g$. As $\alpha$ decreases, $\mathcal{C}_1$ grows but remains the only connected component of  $\lbrace x \, \vert \,  g(x) \geq \alpha\rbrace$, until $\alpha=g(p_2)$ when another connected component $\mathcal{C}_2$ appears. At $\alpha=g(v)$, these two connected components are merged, and we say that the one that appeared last dies and becomes merged with the one that appeared first:  $\mathcal{C}_2$ becomes merged with $\mathcal{C}_1$. Finally, as $\alpha$ gets smaller and goes to $-\infty$, only one connected component, $\mathcal{C}_1$, remains.  
The parameter $\alpha$ can be seen as time (going backwards), thus we speak in terms of birth times, death times, and lifetimes of connected components. In our example $\mathcal{C}_1$ is born at $g(p_1)$ and never dies, and $\mathcal{C}_2$ is born at $g(p_2)$ and dies at $g(v)$. Birth times are greater than death times since time is going backwards, and we define the lifetime of a connected component as its birth time minus its death time. The result of persistent homology is presented in what is called a \textbf{persistence diagram}. Each connected component is represented by a point (birth time, death time) in the plane. More precisely these points are in the half plane, below the diagonal $\lbrace y=x \rbrace$, since birth time is always greater than death time. In our example, we obtain a persistence diagram with two points:  $(g(p_1),-\infty)$ and $(g(p_2),g(v))$, as depicted in Figure \ref{Tomato2} B. Let us point out that a connected component arises at each peak of the function $g$, and that at each valley (local minimum) a connected component dies. With a slight abuse of langage we call $p_i$ the peak of $g$ corresponding to the local maximum $p_i$, and we define $\tau_i$ the \textbf{prominence of a peak} $p_i$, as the lifetime of the connected component that is born at $g(p_i)$. It is important to understand that in practice the result of persistent homology is just a collection of unlabeled (birth time, death time) points in the persistence diagram. The connected component that generated a given point in the diagram is not indicated.\\
  
\begin{figure}[h] 
\begin{center} 
\includegraphics[scale=0.25]{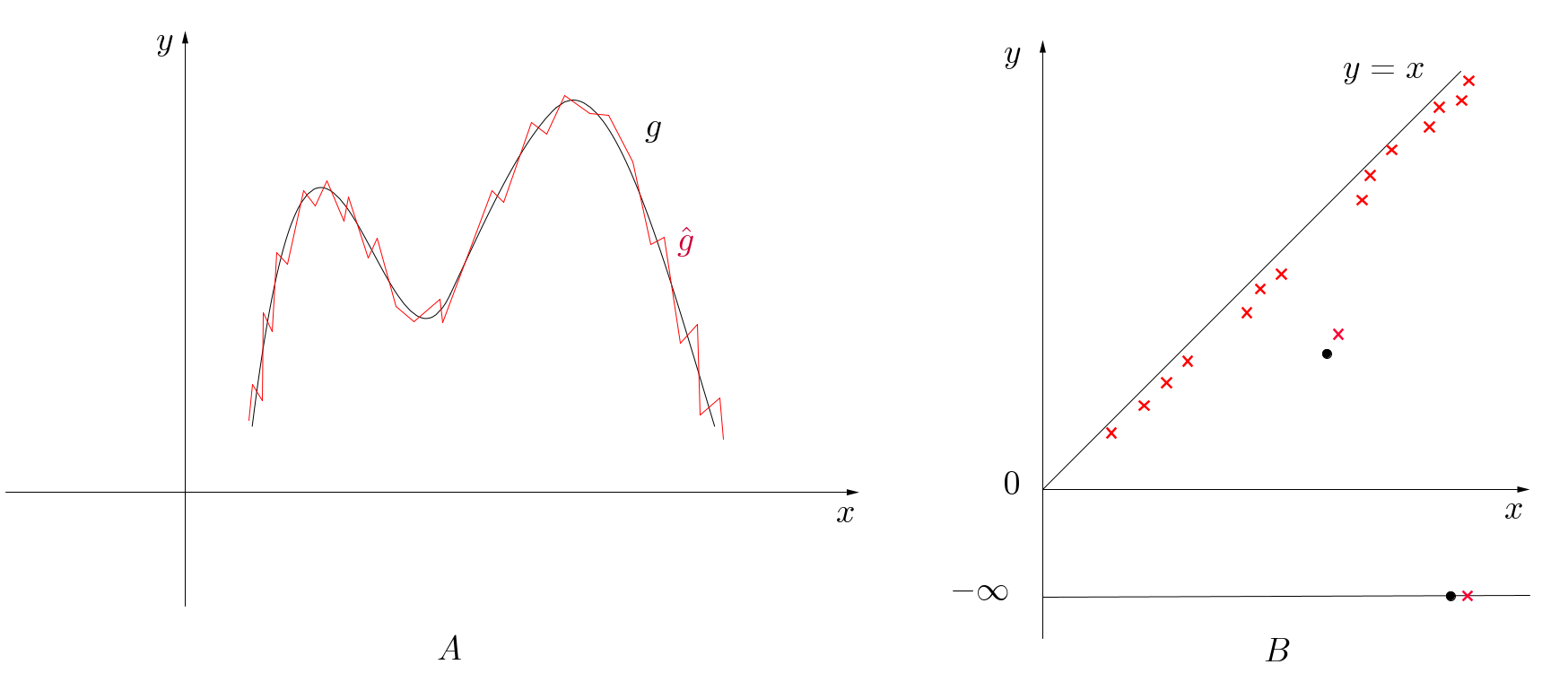}
\end{center}  
\caption{A : A function $g$ and a ``noisy'' approximation $\hat{g}$. B : Their respective persistence diagrams. The persistence diagram of $g$ consists of the $2$ black points, and the persistence diagram of $\hat{g}$ consists of the red crosses.} 
\label{stab} 
\end{figure}

A fundamental property of persistent homology is what is called \textbf{stability}. If a connected component has a short lifetime, this is likely because it corresponds to a small peak of the function, due to noise. Thus the important quantity in a persistence diagram is the points that are far away from the diagonal. If there are a function $g$ and an approximation $\hat{g}$ of $g$ that is close to $g$, then the stability theorem of \cite{PD} states, in simplified terms, that the persistence diagram of $\hat{g}$ and the one of $g$ are also close in the sense that they may differ close to the diagonal, but they should have approximately the same number of points far away from the diagonal. Thus the number of prominent peaks is a stable quantity under perturbations of the function. The stability property is illustrated in Figure \ref{stab}.\\

In persistent homology, all the peaks are merged during the filtration and in the end only one remains\footnote{Actually, if the function is not continuous then its superlevel set of parameter $\alpha$ has several connected components even as $\alpha$ goes to $- \infty$. In this case, in the end several clusters remain.}. ToMATo performs a modified persistent homology, where not all the peaks are merged. The merging condition is based on peak prominence, as explained in Subsection \ref{expl}. 
 
\begin{remark}
Usually in persistent homology, there is a parameter $\alpha$ that increases from $- \infty$ to $+ \infty$ and we are interested in the connected components of the sublevel sets of $g$, rather than superlevel sets with a parameter that decreases. In this setting persistence diagrams are then composed of points above the diagonal and not below. The authors of \cite{Tomato} chose a different perspective that is more natural and appropriate in the context of mode-seeking, but the ideas and results are the same. 
\end{remark} 

\subsubsection{Merging ascending regions based on peak prominence}   \label{expl}
We have recalled the bases of gradient ascent mode-seeking and persistent homology. On one hand with gradient ascent mode-seeking, we have a definition of peaks and their ascending regions, that is, potential clusters. On the other hand with persistent homology, we have a tool to obtain peak prominences leading to a hierarchy of the peaks of $f$. We now see how to combine these two ideas to define a mode-seeking clustering method, that is not sensitive to noise induced peaks of the density estimate.\\    

\begin{figure}
\begin{center} 
\includegraphics[scale=0.3]{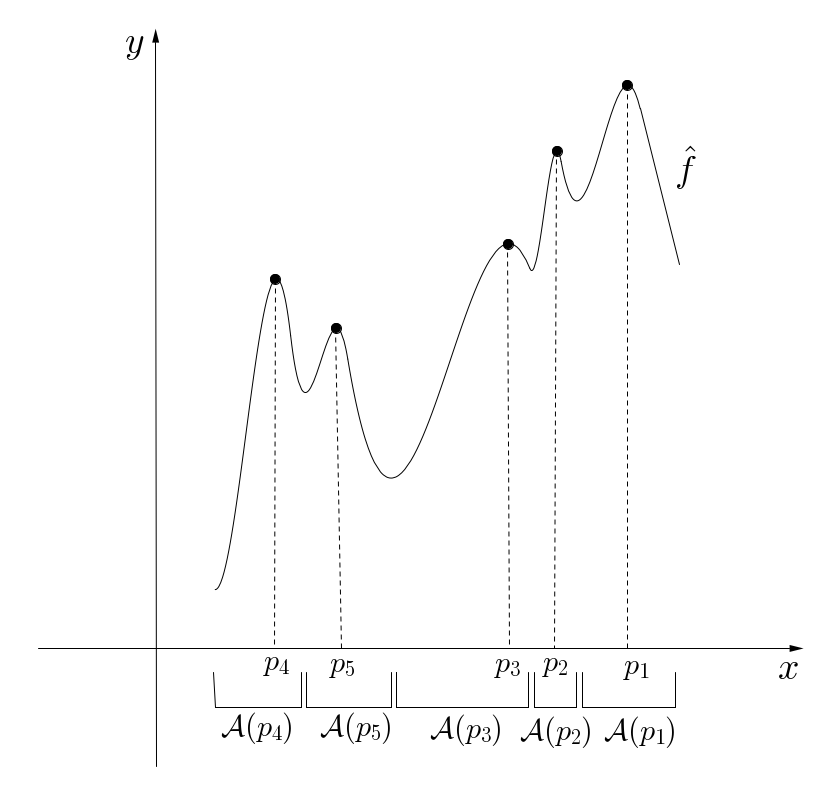}
\end{center}  
\caption{A density estimate $\hat{f}$ with $5$ local maxima $p_1,\dots, p_5$. The ascending regions are denoted by $\mathcal{A}(p_1),\dots \mathcal{A}(p_5)$.} 
\label{Tomato3} 
\end{figure}

Let us work on the example illustrated in Figure \ref{Tomato3}. Here the density estimate $\hat{f}$ of $f$ exhibits $5$ peaks $p_1,\dots, p_5$. The ascending region of a peak $p_i$ is denoted by $\mathcal{A}(p_i)$. A classical mode-seeking algorithm would partition the space according to these ascending regions, resulting in $5$ clusters $\mathcal{A}(p_1),\dots ,\mathcal{A}(p_5)$. However, $\hat{f}$ being an estimate of our data density $f$, visually we would be tempted to think that $p_2,p_3$ and $p_5$ are due to noise, and that the only relevant peaks are $p_1$ and $p_4$, resulting in only $2$ clusters: $\mathcal{A}(p_4) \cup \mathcal{A}(p_5)$ and $\mathcal{A}(p_1) \cup \mathcal{A}(p_2) \cup \mathcal{A}(p_3)$. ToMATo thus departs from the classical mode-seeking setting by allowing merging between ascending regions, taking advantage of persistent homology theory. In this way we are more likely to identify only $2$ clusters, corresponding to unions of ascending regions. \\

To keep only prominent peaks of $\hat{f}$, the idea of ToMATo is to fix a threshold parameter $\tau >0$ and to perform persistent homology but with the additional condition that we merge only peaks of prominence less than $\tau$ with peaks of prominence greater than $\tau$. The choice of parameter $\tau$ is discussed in Section \ref{tau}. It is equivalent to the choice of the number of clusters. \\

In our example we have $ \tau_1 \geq \tau_4 \geq \tau_5 \geq \tau_2 \geq \tau_3 $ and if $\tau$ is chosen such that $\tau_5 \leq \tau \leq \tau_4 $, then:
\begin{itemize}
\item The peak $p_1$ never dies.
\item The peak $p_4$ never dies: it does not become merged with $p_1$ because it has prominence $\tau_4 \geq \tau$.
\item The peaks $p_2$ and $p_3$ are merged with $p_1$.
\item The peak $p_5$ is merged with $p_4$.
\end{itemize}

The final clusters are the unions of ascending regions that got merged together. Here two clusters are obtained: $\mathcal{A}(p_4) \cup \mathcal{A}(p_5)$ and $\mathcal{A}(p_1) \cup \mathcal{A}(p_2) \cup \mathcal{A}(p_3)$, as desired.

\subsubsection{Choosing $\tau$ or equivalently the number of clusters}\label{tau}
In practice, the number of clusters can directly be specified, instead of specifying $\tau$.\\

\begin{figure}[h] 
\begin{center} 
\includegraphics[scale=0.3]{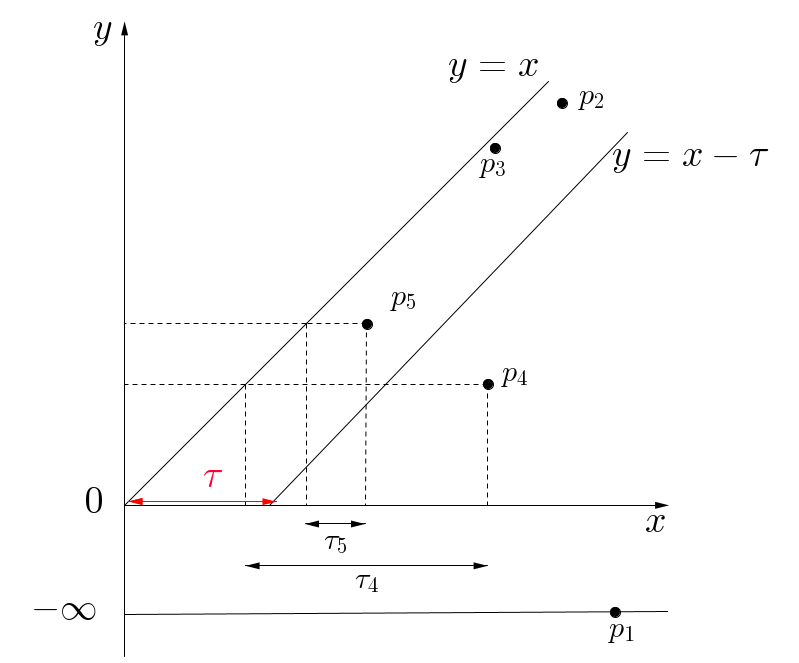}
\end{center}  
\caption{The persistence diagram of the density estimate $\hat{f}$ of Figure \ref{Tomato3}. The parameter $\tau$ is chosen between $\tau_5$ and $\tau_4$ to obtain $2$ final clusters corresponding to the peaks $p_1$ and $p_4$ of the density. Indeed, the persistence diagram points that lie below (respectively above) the line $y=x-\tau$ have lifetime greater (respectively smaller) than $\tau$.
} 
\label{Tomato4}
\end{figure}

The main idea to choose the number of clusters is the following. When performing persistent homology, we obtain the peak prominences and they are represented in the form of a persistence diagram, which in our example would look like the one on Figure \ref{Tomato4}. We recall that in practice the points of the diagram are not labeled. Looking at this persistence diagram, we can see two points far away from the diagonal and other points close to the diagonal, a priori due to noise. We thus decide that we want $2$ clusters. This is equivalent to choosing $\tau_5 < \tau \leq \tau_4 $, as illustrated on Figure \ref{Tomato4}.\\ 

The ToMATo algorithm must thus be run twice. The first time it is run without specifying the number of clusters and it performs classical persistent homology. The result is a persistence diagram from which the number of points far away enough from the diagonal can be chosen as the desired number of clusters. Then, the algorithm is run a second time, specifying the desired number of clusters. \\

We highlight the fact that running the algorithm with $\tau =+ \infty$ is just classical persistent homology: all the clusters that can be merged, are merged. On the opposite, ToMATo with $\tau=0$ does not merge any clusters and the output clusters are the ascending regions of peaks.

\subsection{Explanation of the ToMATo algorithm} \label{algo}
\label{sec:tomato-explanation}

We now explain the ideas of the ToMATo algorithm in the discrete setting from a practical point of view.\\

The inputs to ToMATo are: data points $x_1,\dots,x_n$; a density estimate at these points; the distances between the data points.  Let us denote by $\hat{f}_i$ the value of $\hat{f}$ at $x_i$. The idea of ToMATo is to mimick in the discrete setting the gradient ascent and ascending regions merging previously explained in the continuous setting. The computational cost is reduced by building graphs on top of the data. We therefore need a few definitions about graphs before explaining the algorithm.

\subsubsection{Graph definitions}
Let us recall that a graph is an abstract set of points, called vertices, together with a set of arcs going from one vertex to another vertex, called edges. The edge between a vertex $x_i$ and a vertex $x_j$ is denoted by $e_{ij}$. The vertex $x_i$ is then called the initial vertex of $e_{ij}$, and $x_j$ the final vertex of $e_{ij}$. A graph is said to be undirected if, for all edges $e_{ij}$ we have $e_{ij}=e_{ji}$, otherwise it is said to be directed. Finally, a graph is a binary graph if its edges $e_{ij}$ are equal to either $0$ or $1$. In the following, all graphs are binary. 

\begin{defi} Directed tree (\cite{Koontz}) \label{tree} \\
Let $G$ be a directed graph. 
A set of edges $e_1, \dots, e_m$ is said to be a directed path from a vertex $x$ to a vertex $x'$, if $x$ is the initial vertex of $e_1$, if $x'$ is the final vertex of $e_m$, and if for $k \in \llbracket 1, m-1 \rrbracket$ the final vertex of $e_k$ is the initial vertex of $e_{k+1}$.\\ 
A directed tree is a directed graph with a specified vertex $r$, called its root, such that:
\begin{enumerate}
\item Every vertex $x \neq r$ is the initial vertex of exactly one edge.
\item The vertex $r$ is the initial vertex of no edge.
\item There is no directed path from a vertex to itself (i.e. no cycles).
\end{enumerate}
\end{defi} 

\begin{figure}
\centering
\includegraphics[scale=0.3]{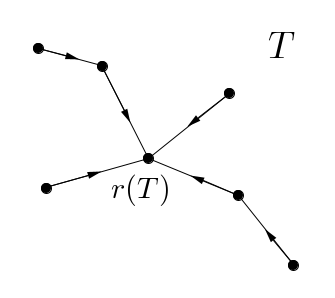}
\caption{An example of a directed tree $T$, with its root $r(T)$.}
\label{arbre}
\end{figure}

An example of a directed tree is depicted in Figure \ref{arbre}. Directed trees will be, in the discrete setting, the equivalent of the ascending regions of the continuous setting.   

\begin{defi} Upper star \\
Let $G$ be a graph, and assume that there is a function defined on the vertices of $G$. Let us denote by $x_1, \dots, x_m$ the vertices of $G$. For $k \in \llbracket 1, m \rrbracket$, we define the upper star of the vertex $x_k$ in $G$, as the set of edges connecting $x_k$ to other vertices with higher function values, along with these vertices. The set of vertices of the upper star of $x_k$ is denoted by $S_k$.
\end{defi} 

\begin{Ex} \label{ex_section}
Constructing the $S_k$ sets for Fig.~\ref{graph} we get: $S_1=\emptyset$, $S_2=\{1\}$, $S_3=\{1\}$, $S_4=\emptyset$, $S_5=\{1\}$, $S_6=\{1\}$, $S_7=\{4,6\}$, , $S_8=\{4\}$, $S_9=\{4\}$, $S_{10} = \emptyset$, $S_{11} = \{10\}$, $S_{12} = \emptyset$, $S_{13} = \{10,12\}$, $S_{14} = \{12\}$, $S_{15} = \{12\}$, $S_{16} = \{10\}$, $S_{17} = \{1\}$, $S_{18} = \{10\}$, $S_{19} = \{1,10\}$, $S_{20}=\{ 1\}$.
\end{Ex}

\subsubsection{Principle of ToMATo}
\label{sec:tomato-principle}
We are now ready to explain the pseudo-algorithm of ToMATo. In a preprocessing step, the ToMATo algorithm computes a neighbourhood graph $G$ of the data, which is typically a Rips graph or a $k$-nearest neighbour graph. The Rips graph with parameter $r >0$ is the (undirected) graph where two points are connected if and only if the distance between them is less than $r$. A k-nearest neighbour graph is a graph where each point is connected to its k-nearest neighbours. The choice of $r$ or of $k$ is important.
\begin{figure}  
\centering  
\includegraphics[scale=0.15]{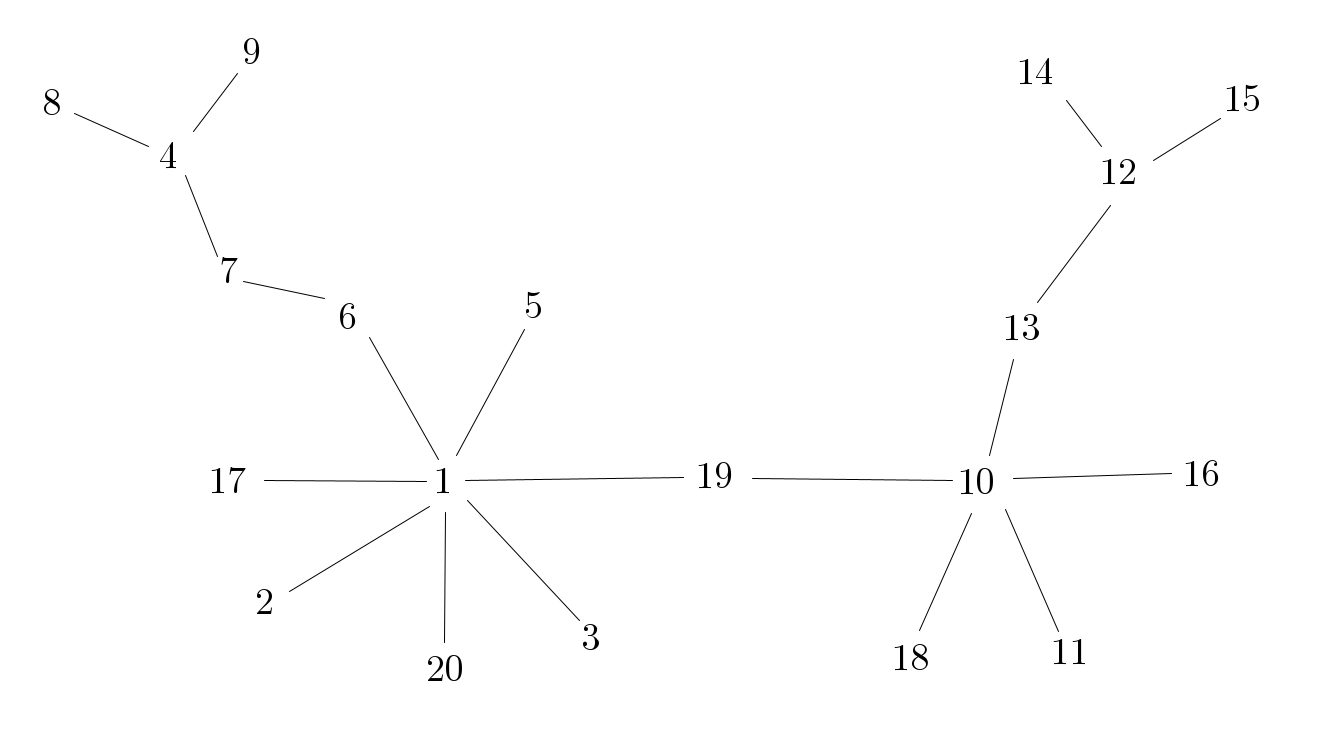}
\caption{A well-chosen neighbourhood graph of a dataset of points in $\mathbb{R}^2$. The number $i$ corresponds to a point $x_i$, where the indexes have been reordered such that $\hat{f}_1 \geq \dots \geq \hat{f}_n$. 
This is not completely realistic since a neighbourhood graph would naturally have more edges than the ones drawn.}
\label{graph} 
\end{figure}
In Figure \ref{graph} a well-chosen neighbourhood graph of a dataset of points in $\mathbb{R}^2$ is represented as an illustration. \\ 

Let $G$ be a neighbourhood graph build on top of the data points. The principle of the algorithm is the following: 
\begin{itemize}
\item Build the directed trees in $G$ having root at a local maximum of $\hat{f}$.
\item Potentially merge some trees to obtain the desired number of clusters.
\end{itemize}

We emphasize the parallel with the continuous setting. Building directed trees having root at a local maximum of $\hat{f}$ is a mode-seeking step, it is the equivalent of searching peaks and their ascending regions. In the continuous case ascending regions are defined thanks to a gradient ascent, but there is no gradient needed here in the discrete setting for the tree construction. Each tree is then a potential cluster, and allowing merging between trees enables to recover some stability. The output of the algorithm is a union of directed trees, each union corresponding to a final cluster of the set of data points. \\

\begin{figure} 
\centering
\includegraphics[scale=0.15]{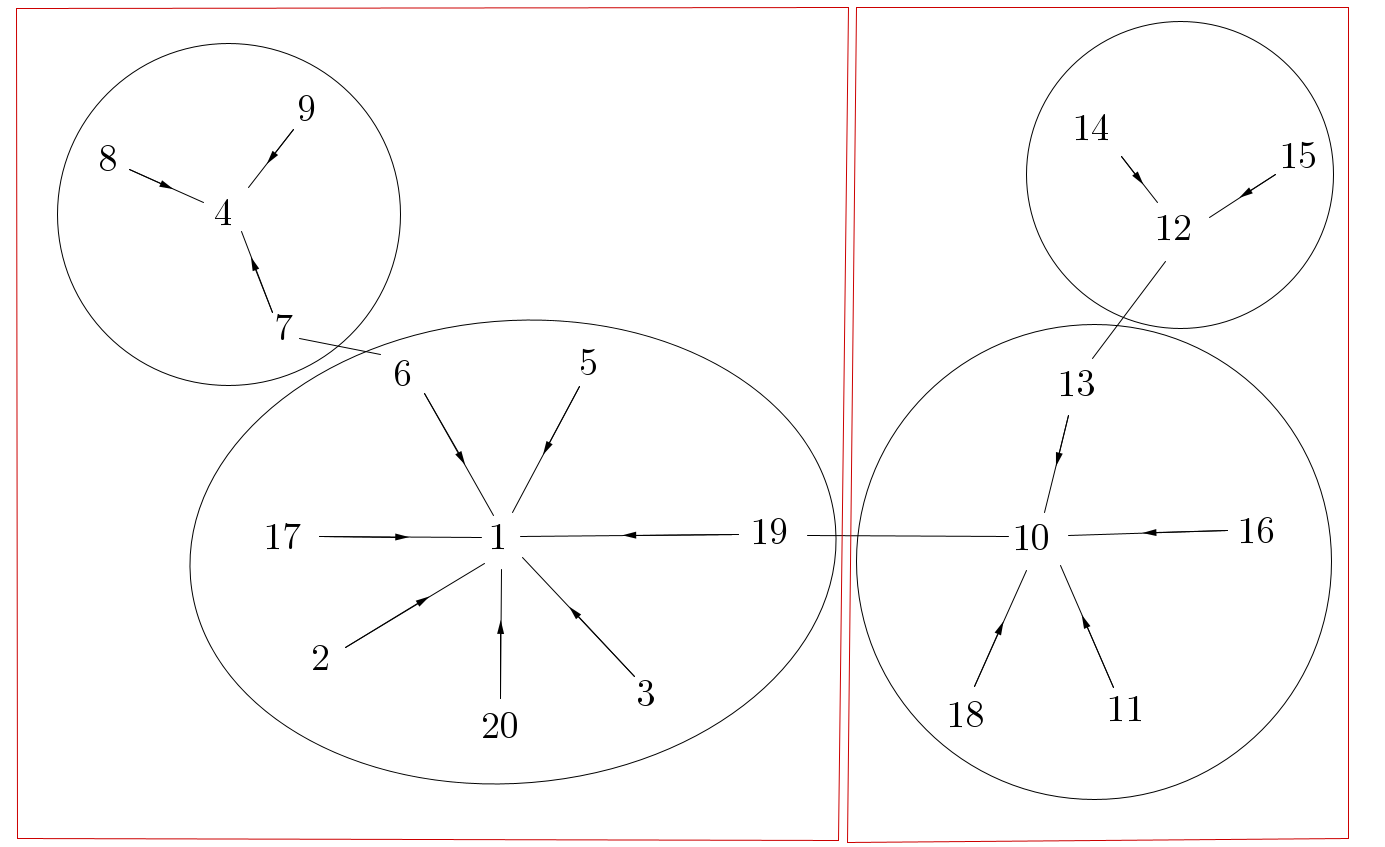}
\caption{ $\left[  \textrm{Continuation of Figure \ref{graph} and Example \ref{ex_section}.}\right]$ 
The result of the ToMATo algorithm with $\tau$ chosen such that there are $2$ final clusters. Each tree is circled in black, and the unions of trees circled in red correspond to the $2$ final clusters. 
}
\label{graph_trees} 
\end{figure}

We explain more precisely the tree construction. A vertex $x_i$ is declared a peak if the set $S_i$ is empty. If a vertex $x_i$ is not a peak then it is attached to the tree containing $\textrm{argmax}_{j \in S_i} \, \hat{f}_j$, forming a cluster. Iterating over all vertices, directed trees are thus created, the roots of which are peaks of $\hat{f}$ in the neighbourhood graph. These trees constitute the ``ascending regions". In Figure \ref{graph_trees}, each directed tree computed by the algorithm for the dataset of Figure \ref{graph} is circled. They are the potential clusters without any merging. 
\\

We now explain how to perform persistent homology on the neighbourhood graph. We directly explain the modified persistent homology with the condition that for a paramater $\tau$, we merge only peaks of prominence less than $\tau$ with peaks of prominence greater than $\tau$. Classical persistent homology is the same procedure, with $\tau=+\infty$. For each vertex $x_i$, we denote by $T_i$ the (necessarily unique) tree which includes $i$, and $r(T_i)$ the root of $T_i$. If two vertices $x_i$ and $x_k$ are in the same tree, we denote this tree by either $T_i$ or $T_k$; for instance we always have $T_i=T_{r(T_i)}$.
The algorithm is the following: let $\tau >0$, let $x_i$ be a vertex,\\
\texttt{for $j$ in $S_i$: (\textit{we check if other trees $T_j$ can be merged with $T_i$})\\  
\smallskip 
\hspace{20pt} if $\hat{f}_{r(T_j)}-\hat{f}_i<\tau$: merge $T_j$ with $T_i$ \\ 
Let $T_{\max}(i)$ be, among the trees intersecting $S_i$ having a higher root than $T_i$, the tree with the highest root: \\
\smallskip  
if $\hat{f}_{r(T_i)}-\hat{f}_i<\tau$: merge $T_i$ with $T_{\max}(i)$ \\ (\textit{we check if $T_i$ itself can be merged with another tree}).} \\
 
In full words it means that if $x_i$ connects two peaks $T_i$ and $T_j$, respectively $T_i$ and $T_{\max}(i)$, then it is near a local minimum of $\hat{f}$. The prominence of the smallest peak is thus estimated respectively by $\hat{f}_{r(T_j)}-\hat{f}_i$ or $\hat{f}_{r(T_i)}-\hat{f}_i$, and if this estimated prominence is smaller than $\tau$  
then the smallest peak is merged with the highest peak. 

\begin{Ex} $\left[ \textrm{Continuation of example in Figures \ref{graph} and \ref{graph_trees}}. \right]$ \\
Let us illustrate the algorithm for different values of $i$.
\begin{itemize}
\item Let $i=13$:\\
$S_{13}=\lbrace 12,10 \rbrace$, thus $x_{13}$ is attached to the tree $T_{10}$.\\
$j=12$: if $\hat{f}_{r(T_{12})}-\hat{f}_{13}=\hat{f}_{12}-\hat{f}_{13}<\tau$ then $T_{12}$ is merged with $T_{13}=T_{10}$.\\
$T_{\max}(13) =\emptyset$ (thus no further merging can be done).  
\item Let $i=7$:\\
$S_7=\lbrace 6,4 \rbrace$, thus $x_{7}$ is attached to the tree $T_{4}$.\\
$j=6$: if $\hat{f}_{r(T_6)}-\hat{f}_{7}=\hat{f}_{1}-\hat{f}_{7}<\tau$ then $T_{6}=T_{1}$ is merged with $T_{7}=T_{4}$.\\
$T_{\max}(7) =T_1$. \\
If $\hat{f}_{r(T_7)}-\hat{f}_7=\hat{f}_{4}-\hat{f}_7<\tau$ then $T_7=T_4$ is merged with $T_1$. 

\end{itemize}

\end{Ex}

If $\tau=+\infty$, the above procedure is the persistent homology of the filtration formed by the neighbourhood graphs of $\lbrace x_i \, \vert \,  \hat{f}_i \geq \alpha \rbrace$, for $+ \infty \geq \alpha \geq -\infty$.  In Figure \ref{graph_trees}, we circle in red the $2$ final clusters that we obtain when applying ToMATo with a value of $\tau$ chosen such that we obtain $2$ clusters. 

\begin{remark}
In the above procedure, trees can in fact be unions of trees.     
\end{remark}

\begin{remark}
If the neighbourhood graph has several connected components, then they are never merged during persistent homology. Consequently, when ToMATo is run with $\tau=+\infty$ and performs persistent homology of the graph, there are as many points at $y=-\infty$ in the diagram, as there are connected components in the graph. The number of connected components is thus also the minimal number of clusters that ToMATo can find, with any value of $\tau$. 
\end{remark}

We have thus defined in the discrete setting, relevant notions of peak, peak prominence, ascending region, and persistent homology. Therefore everything that we explained in the continuous setting adapts to the discrete setting. \\
  
We do not go into the details of the ToMATo implementation. The actual implementation is really clever and optimized, and a bit different from the pseudo-code presented in \cite{Tomato} and that is explained here. We only presented the general ideas, that are sufficient to understand the algorithm and its most important parameters, the latter being detailed in Subsection \ref{param}.

\begin{remark} A few words about theoretical guarantees \\
The authors of \cite{Tomato} proved that the algorithm recovers the exact number of clusters, and that the output clusters coincide with the peaks of the density $f$. Let us present the results in a simplified way. For results $1$ and $2$, let us assume that the $n$ data points  are i.i.d. and sampled under a density $f$ that is known.\\

Result 1 (\cite{Tomato}, theorem 4.8): \\
If the persistence diagram of $f$ has a signifant gap between points close to the diagonal and points far away from the diagonal, and if $n$ is large enough, then there exist a parameter $r$ for the Rips graph and a parameter $\tau$ such that with high probability, the number of clusters computed by the algorithm is equal to the number of peaks of $f$ of prominence greater than $\tau$. \\

Result 2 (\cite{Tomato}, theorem 4.9): \\
If the persistence diagram of $f$ has a signifant gap between points close to the diagonal and points far away from the diagonal, and if $n$ is large enough, then there exist a parameter $r$ for the Rips graph and a parameter $\tau$ such that with high probability, for each peak of $f$ of prominence greater than $\tau$, the algorithm outputs a cluster that coincides with the part of the ascending region that correpond to the top of the peak. \\

A third result states that with an approximation $\hat{f}$ of $f$ that is close to $f$, results 1 and 2 still hold. More precisely:\\

Result 3 (\cite{Tomato}, section 5): \\
If the persistence diagram of $f$ has a signifant gap between points close to the diagonal and points far away from the diagonal, if $n$ is large enough, and if $\hat{f}$ is close enough to $f$, then there exist a parameter $r$ for the Rips graph and a parameter $\tau$ such that with high probability:
\begin{itemize}
\item The number of clusters computed by the algorithm with input $\hat{f}$, is equal to the number of peaks of $f$ of prominence greater than $\tau$.
\item For each peak of $f$ of prominence greater than $\tau$, the algorithm with input $\hat{f}$, outputs a cluster that coincides with the part of the ascending region that correpond to the top of the peak. 
\end{itemize}                                    
\end{remark}

\subsection{ToMATo in practice with Gudhi} \label{param}
ToMATo is implemented in Gudhi\footnote{\url{https://gudhi.inria.fr/python/latest/clustering.html}.}, and well documented. The ToMATo tutorial\footnote{\url{https://github.com/xetaxe/ToMATo-Notebook/blob/master/Guide\%20ToMATo.ipynb}.} can also be consulted. The most important parameters are: 
\begin{itemize}
\item `density\_type': the density type  can be chosen among `KDE', `$\log$ KDE', `DTM' or `$\log$ DTM'. The DTM (Distance To a Measure) can be seen as an improvement over the k-NN estimator and is defined in \cite{DTM_density}. The ToMATo parameters `k\_DTM',`q' and `dim' are associated with the DTM\footnote{In high dimension one may need to change the `dim' parameter. This is explained in the Gudhi documentation.}, see \cite{DTM_density} for details.  
\item `graph\_type': the graph type can be chosen among  `k\_nn' or `radius'. The choice `radius' is the Rips graph ; the parameter $r$ of which is to be specified by the user. The choice `k\_nn' is the k-nearest neighbour graph, with default $k=10$. 
\item `n\_clusters': the number of clusters.
\end{itemize}          
  
The default parameter values are `density\_type=$\log$ DTM' and `graph\_type=k\_nn' with $k=10$. In practice the algorithm has to be run first without specifying `n\_clusters', giving a persistence diagram. Then we see how many points are sufficiently far away from the diagonal, and the algorithm has to  be rerun with this number as `n\_clusters'. \\

In our experience, the values that give the best results are usually the default values. By ``best results", we mean that the persistence diagram presents a clear gap between points close to the diagonal and points far away from the diagonal, allowing an unequivocal choice of the number of clusters. ToMATo thus requires tuning only a few parameters.
  
\section{Application of ToMATo in a spike sorting problem} \label{app}

The problem of spike sorting is the following. The raw data consists of the recording of mixed activity of multiple neurons, and the goal is to recover how many neurons contributed to the recording, and find the times at which each neuron fired a spike. The result of spike sorting is called a rasterplot. A rasterplot is a collection of discrete time series, presented in rows. Each row is a time series that corresponds to an identified neuron, and it shows the series of the times at which this neuron emitted a spike. Each discrete time series is represented as a series of bars. A rasterplot is showed in Figure \ref{dgm_raster} in Subsection \ref{appli} when ToMATo is performed on real data.\\   

We apply ToMATo in the context of spike sorting on both simulated and real data, and we explain how it supersedes the usual method.    

\subsection{General outline and interest of the ToMATo method}
\label{sec:spike-sorting-short}
We explain the general spike sorting outline. \\

Neuronal activity is recorded at different sites. In our simulated and real data there are $4$ sites. We make the assumption that anytime a neuron ﬁres a spike, the same underlying waveform with some additive auto-correlated Gaussian noise with variance $1$ is recorded on each site. More precisely, there is one waveform per electrode and per neuron. The raw data, or the recorded neuronal activity, is then the mixture of the activity of several neurons.\\

First a preprocessing step to detect spikes and spike times, is performed. The goal is to identify times at which the recording presents a large local maximum at at least one of the $4$ sites: the waveforms around such times are spike candidates. A collection of $4$ waveforms corresponding to the detected spikes of a neuron on the $4$ sites, is called an event. Each waveform on each site has $45$ sampling points, thus each event is a $4 \times 45=180$ dimensional vector. At this point, this is a clustering problem. The goal is to cluster the events, each obtained cluster being interpreted as the spikes from one neuron.\\

After clustering, the final step is to go back to the raw data and to the spike times, and assign each spike time to a neuron in order to obtain a rasterplot. \\

In the usual spike sorting procedure, clustering cannot be performed directly. The clustering phase,  explained in full details in \cite{SpySort} and reexplained in Appendix \ref{annexe}, can be summarized with the following steps:
\begin{enumerate}
\item \textbf{Obtaining clean events (eliminating superpositions)}:  \\ 
A superposition occurs when two different neurons fire nearly simultaneously, leading to an event that is the superpersition of the spikes of the two neurons. In a first step, the most obvious superpositions are eliminated, to keep only what are called ``clean" events\footnote{When the model will be subsequently used to classify data, superpositions have to be looked for and accounted for.}. This step is essential for the principal component analysis that follows, otherwise superpositions skew the principal components. 
\item \textbf{Dimension reduction with principal component analysis}: \\
A principal component analysis is performed on the clean events, to reduce the dimension from $180$ to a smaller integer $d$, typically $d \leq 10$. 
\item \textbf{Dynamic visualization of the data to find the number of clusters}: \\
Projections on the principal components of the data are visualized to help the user choose the number of clusters. This step is not reproductible, the choice of the number of clusters depends a lot on the user.
\item \textbf{Automatic clustering}:\\
An automatic clustering method in $\mathbb{R}^d$ is applied on the projections on the principal components of the data,  with K clusters where the integer K has been chosen in the previous step. K-means is often used since it is a common method that gives satisfying results, however in the case where clusters seem to be non convex in the visualization step, other more appropriate clustering methods are used.
\end{enumerate}    

This procedure is laborious. Let us explain how ToMATo supersedes these four steps.\\
 
A first run of ToMATo, i.e., without specifying `n\_clusters', provides an efficient way to choose the right number of clusters, by looking at the persistence diagram. Thus, the visualization of step $3$ is no longer necessary. Since dimension reduction is performed mostly to enable data visualization, and since the superposition elimination is performed to not skew the dimension reduction, step $1$ and $2$ become superfluous as well. Finally, once the number of clusters is chosen, a second run of ToMATo is an automatic clustering method, the equivalent to step $4$. ToMATo alone thus replaces the whole four-step procedure. Let us highlight some facts:
\begin{itemize}
\item For both simulated and real data, we apply ToMATo directly in high dimension ($180$) and it turns out that it works and runs within a reasonable time. In fact, with a dimension reduction it can be seen that our data intrinsically live at most in a $5$ or $6$ dimensional space, explaining why the high dimension does not raise an issue. For general data however, it may  be necessary to perform a dimension reduction before applying ToMATo. 
\item The persistence diagrams obtained with ToMATo on our data often allow an unequivocal choice of the number of clusters, but sometimes there can still be an ambiguity. This is discussed in Subsection \ref{simu}.
\item Our data present two main perturbation sources. The first one consists of the presence of superpositions, and the second one consists of small waveforms. In the simulations in Subsection \ref{simu} we find that ToMATo is extremely robust to superpositions. This fact shows that the superposition elimination of the usual spike sorting procedure becomes indeed unnecessary with ToMATo.
\end{itemize} 

\subsection{Simulations}\label{simu}

\begin{figure} 
\centering  
\includegraphics[scale=0.3]{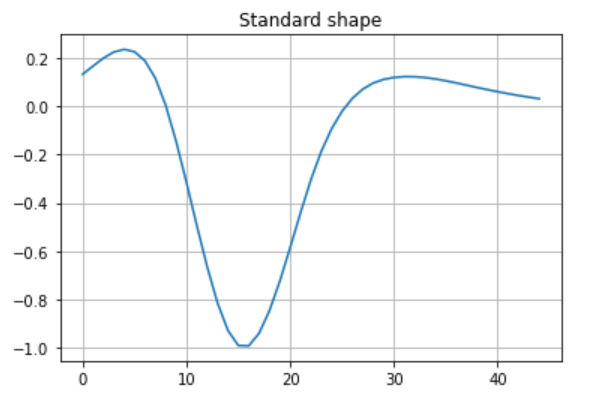}
\caption{The chosen standard shape.}
\label{mothershape}
\end{figure}

We simulate directly the events on the $4$ sites. To do so, we consider a standard spike shape that we multiply by a scale factor, called hereafter amplitude, to obtain the ideal waveform for each neuron at each site. The considered standard shape is displayed in Figure \ref{mothershape}. Each neuron is thus characterized by $4$ amplitudes. Then, some Gaussian noise is added (with constant variance equal to $1$), to simulate the events. Let us recall that each event has $4 \times 45=180$ sampling points, it is thus a $180$ dimensional vector.
Finally, to create a superposition, $2$ neurons are randomly selected, then the event of the second neuron is shifted from the event of the first one with a small random interval, such that the spikes of the two neurons are almost aligned. A superposition thus generates a different shape.\\

\newpage

The three main parameters for the simulations are:
\begin{enumerate}
\item the number of neurons,
\item the amplitudes,
\item the superposition frequence.
\end{enumerate}

In the rest of Subsection \ref{simu}, we apply ToMATo with all parameters set to default, except the `dim' parameter that we set to $2$ because of the high dimension. 

\subsubsection{A first example} 
\begin{figure}  
\centering  
\includegraphics[scale=0.3]{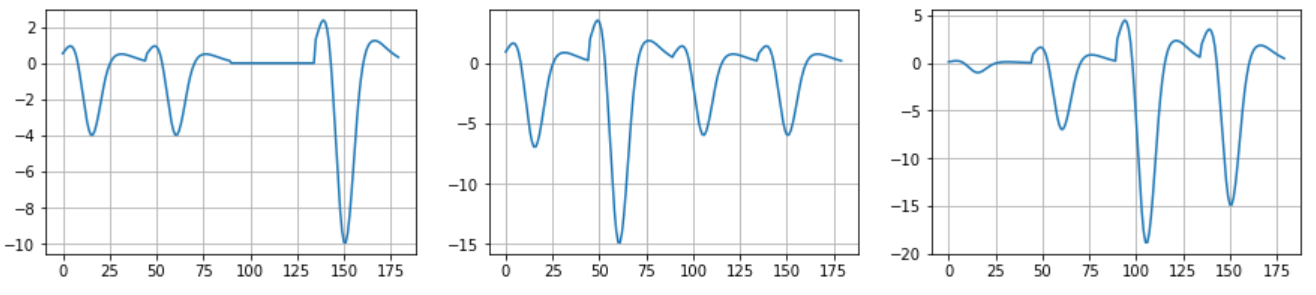}
\caption{The $3$ ideal shapes. Each shape corresponds to a neuron. For each shape, the $4$ ideal waveforms on the $4$ sites are concatenated horizontally.}
\label{ideal}
\end{figure}

\begin{figure} 
\centering  
\includegraphics[scale=0.35]{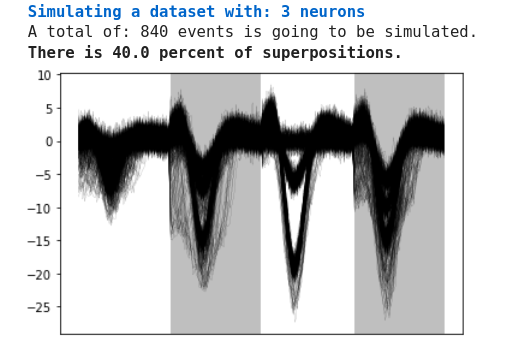}
\caption{Simulated events.}
\label{evts}
\end{figure}

To illustrate the principle of our simulations we present a first example. We consider $3$ neurons, and amplitudes between $0$ and $20$. The $3$ ideal shapes that we use, are depicted in Figure \ref{ideal}. We simulate events by adding Gaussian noise with constant variance equal to $1$, and we add $40$ percent of superpositions. Simulated events are displayed in Figure \ref{evts}.\\

\begin{figure}
\centering  
\includegraphics[scale=0.25]{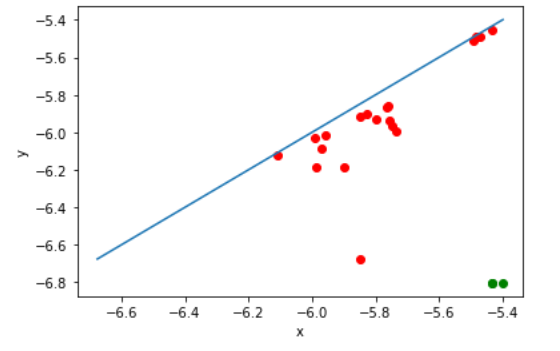}
\includegraphics[scale=0.26]{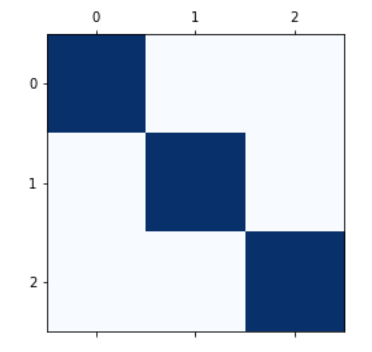}
\caption{Left: Persistence diagram. Right: Confusion matrix. The color scale goes from light blue ($0$) to dark blue ($1$).}  
\label{dgm&conf}
\end{figure}

\begin{figure} 
\centering  
\includegraphics[scale=0.3]{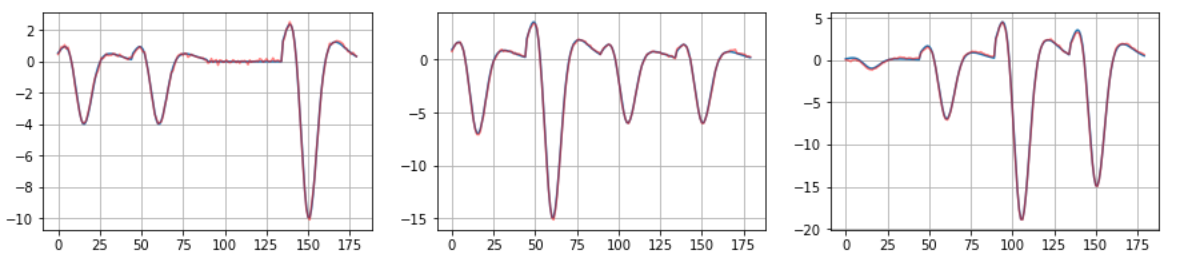}
\caption{Event medians of identified clusters (in red), superposed with the ideal shapes (in blue).}
\label{medians}
\end{figure}

We apply ToMATo on this event collection. We first apply ToMATo without specifying `n\_clusters' to obtain the persistence diagram that is depicted in Figure \ref{dgm&conf} Left. Green points are points with ordinate $y=-\infty$. There are unequivocally $3$ points far away from the diagonal. We then rerun ToMATo with `n\_clusters'$=3$. To assess the quality of clustering, we compute the event medians of identified clusters and superpose them with the ideal shapes. This is illustrated in Figure \ref{medians}. We can see that the medians and the ideal shapes superpose perfectly. We also plot a confusion matrix (Figure \ref{dgm&conf} Right): the entry $(i,j)$ corresponds to the percentage of events of neuron $i$ assigned to cluster $j$. In the rest of this Subsection, only confusion matrices are displayed to summarize clustering results. \\

\subsubsection{On the automatic identification of the number of neurons}
\label{sec:number-of-neurons}
\begin{figure} 
\centering  
\includegraphics[scale=0.3]{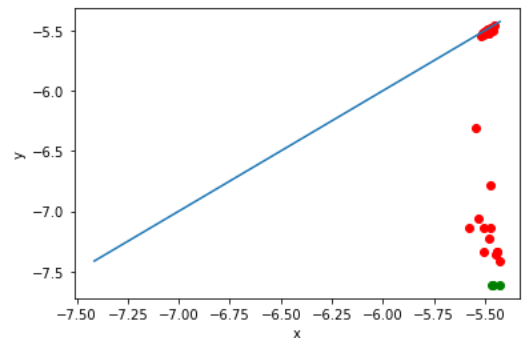}
\includegraphics[scale=0.3]{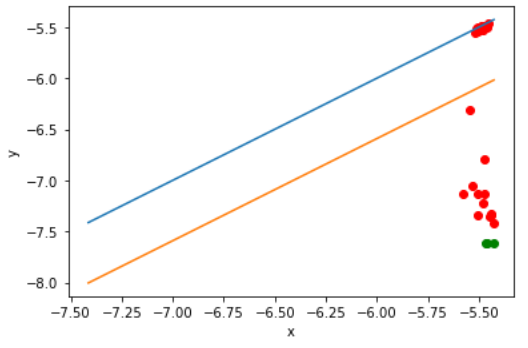}
\caption{Left: A persistence diagram obtained from a simulation with $15$ neurons, a superposition frequence of $0.05$, amplitudes between $0$ and $20$. Right: The same diagram, with, in orange, the line with equation $y=x- \tau$ where $\tau$ is the mean of the lifetimes of the points. We detect $15$ points below this line.}
\label{dgm_detect2}
\end{figure}

\begin{figure} 
\centering  
\includegraphics[scale=0.3]{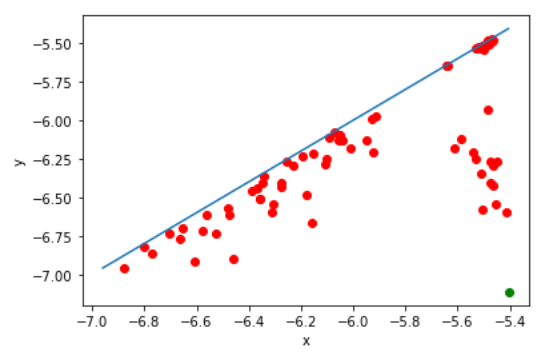}
\includegraphics[scale=0.3]{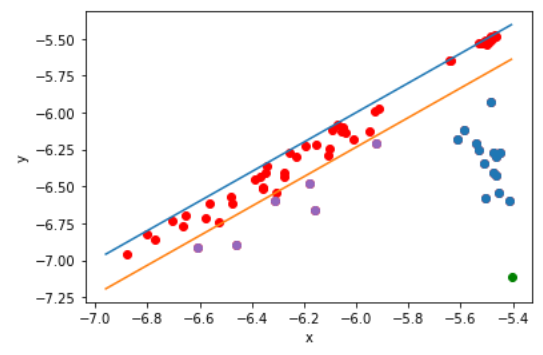}
\caption{Left: A persistence diagram obtained from a simulation with $15$ neurons, a superposition frequence of $0.9$, amplitudes between $0$ and $20$. Right: The same diagram, with, in orange, the line with equation $y=x- \tau$ where $\tau$ is the mean of the lifetimes of the points. Points below this line are cut in two groups, blue and purple. The blue group is the rightmost group and we detect $15$ points in this group.}
\label{dgm_detect1}
\end{figure}

Let us say a few words about the detection of the number of clusters on the persistence diagram. We implemented an automatic identification method adapted to our type of data. After performing simulations in the setting of Subsubsection \ref{sim_res}, it turns out that we mostly obtain two types of diagrams. When the perturbation level (from either superpositions, or low amplitudes or both) is small, persistence diagrams look like the one in Figure \ref{dgm_detect2} Left. They present a clear gap between an agglomeration of points extremely close to the diagonal, and points far away from the diagonal. When the perturbation level is important, persistence diagrams tend to look like the one in Figure \ref{dgm_detect1} Left. This diagram shows one group of points on the top right of the diagram. They are not exactly points far away from the diagonal, but our experiments showed that in this case the right number of neurons is the number of points in this group. \\
 
To detect the number of neurons, we first draw a parallel line to the diagonal, with equation $y=x- \tau$ where $\tau$ is the mean of the lifetimes\footnote{We keep the persistent homology vocabulary. The ``lifetime" of a point in a persistence diagram is an abuse of langage to talk about the lifetime of the peak corresponding to this point. The lifetime of a point $(x,y)$ is thus defined as $(x-y)$.} of the points. It gives a first threshold to detect points far away from the diagonal. If the points below this line show only one group of points, we count the points in this group and take it to be the number of neurons: this is what happens in Figure \ref{dgm_detect2} Right. If the points below this line show two groups of points, we identify these two groups by performing a K-means with $K=2$, on the birth times\footnote{This is still a persistent homology vocabulary. The birth time of a point $(x,y)$ is its absissa $x$.} of the points. Then we take the number of points in the rightmost group, as the number of clusters. This is what happens in Figure \ref{dgm_detect1} Right, where we take only the blue points. It turns out that the diagrams obtained with our type of data always look like Figure \ref{dgm_detect2} or Figure \ref{dgm_detect1}, so that this detection method works well and finds the right number of clusters when possible. All the simulations in Subsection \ref{sim_res} have been performed with this identification method. 

\subsubsection{Simulation results}\label{sim_res}

\paragraph{Realistic simulation of locust and hippocampus data}$ $\\

Our simulations are motivated by real life data. We focus particularly on two cases of interest: the simulation of locust data and the simulation of hippocampus data. Locust data is characterized by the following set of parameters:
\begin{enumerate}
\item There are between $5$ and $15$ neurons.
\item The superposition frequence is smaller than $0.1$, typically around $0.03$.
\item Amplitudes are between $0$ and $20$.
\end{enumerate}
This is a priori an easy setting: there are not too many neurons, not too many superpositions, and amplitudes are high. Our second study case is hippocampus data, which is more complicated. There can be up to $30$ neurons, amplitudes are small and there are a lot of superpositions. More precisely for hippocampus data, we have that:
\begin{enumerate}
\item There are between $15$ and $30$ neurons.
\item The superposition frequence is between $0.1$ and $0.5$, typically around $0.3$.
\item Amplitudes are between $0$ and $10$.
\end{enumerate}
For both cases we simulate $50, 100$ or $200$ events per neuron. We apply the ToMATo method and we present a few results, that are typical of what we generally obtain. \\

\begin{figure} 
\centering     
\includegraphics[scale=0.3]{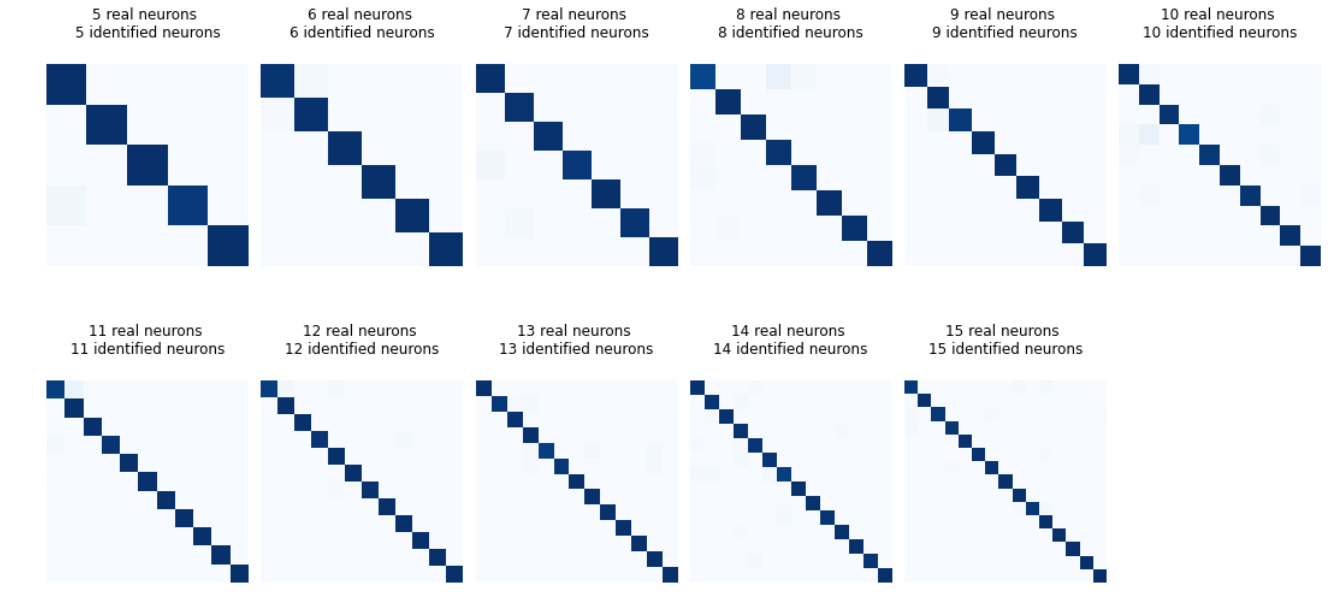}
\caption{Confusion matrices for simulated locust data. $3$ percent of superpositions, $0 \leq $ amplitudes $\leq 20$.}
\label{locust}
\end{figure}

First we show in Figure \ref{locust} the simulation results for locust data. The superposition frequence is set at $0.03$, and the number of neurons vary from $5$ to $15$. The obtained confusion matrices have most of the weight on the diagonal: our clustering results are thus almost perfect.\\

\begin{figure} 
\centering      
\includegraphics[scale=0.3]{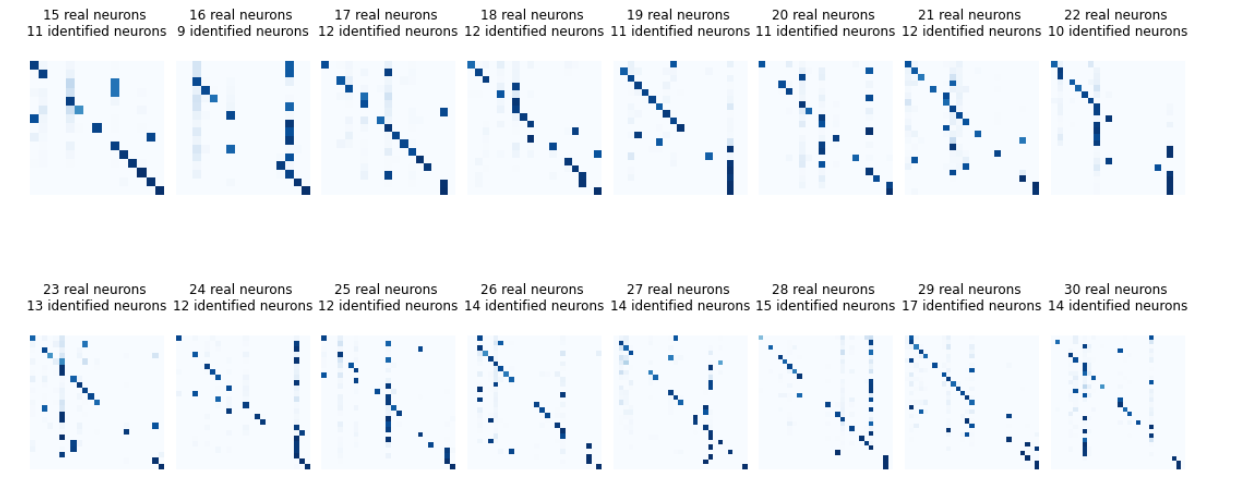}
\caption{Confusion matrices for simulated hippocampus data. $30$ percent of superpositions, $0 \leq $ amplitudes $\leq 10$.}
\label{hippo}
\end{figure}

\begin{figure} 
\centering   
\includegraphics[scale=0.3]{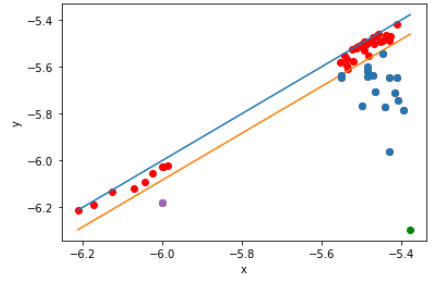} 
\caption{The persistence diagram for simulated hippocampus data of Figure \ref{hippo}, in the case of $29$ neurons. $17$ neurons are identified (points in the blue group). } 
\label{dgm_hippo} 
\end{figure}

In Figure \ref{hippo} we show the results for the simulation of hippocampus data. The superposition frequence is set at $0.3$, and the number of neurons vary from $15$ to $30$. For this type of data we do not recover all the neurons. This was predictable since it is always a difficult type of data to work with. In Figure \ref{dgm_hippo} we plot one of the persistence diagrams, to illustrate the effect of perturbations on the identification of the number of neurons. We can see that there is no clear separation between points close to and far away from the diagonal. This is due to the small amplitudes. However, in this difficult context, identifying about $15$ over $30$ neurons is satisfying, especially if the neurons that are identified, are properly identified. As we can see in the confusion matrices, this is the case. \\ 

To conclude, ToMATo works perfectly for locust data, and it gives the most satisfying results as possible, for hippocampus data.

\paragraph{Robustness of ToMATo to perturbations} $ $\\

It is interesting to explore other parameter values even if they do not reflect a real life setting. In particular, it is interesting to study the robustness of ToMATo to superpositions and to low amplitudes since these are the two main perturbation sources in experimental (real) data. We present some results here. For a fixed number of neurons, we simulate events with different superposition frequences from $0.01$ to $1$, once for amplitudes between $0$ and $20$ and once for amplitudes between $0$ and $10$. \\

\begin{figure} 
\centering      
\includegraphics[scale=0.3]{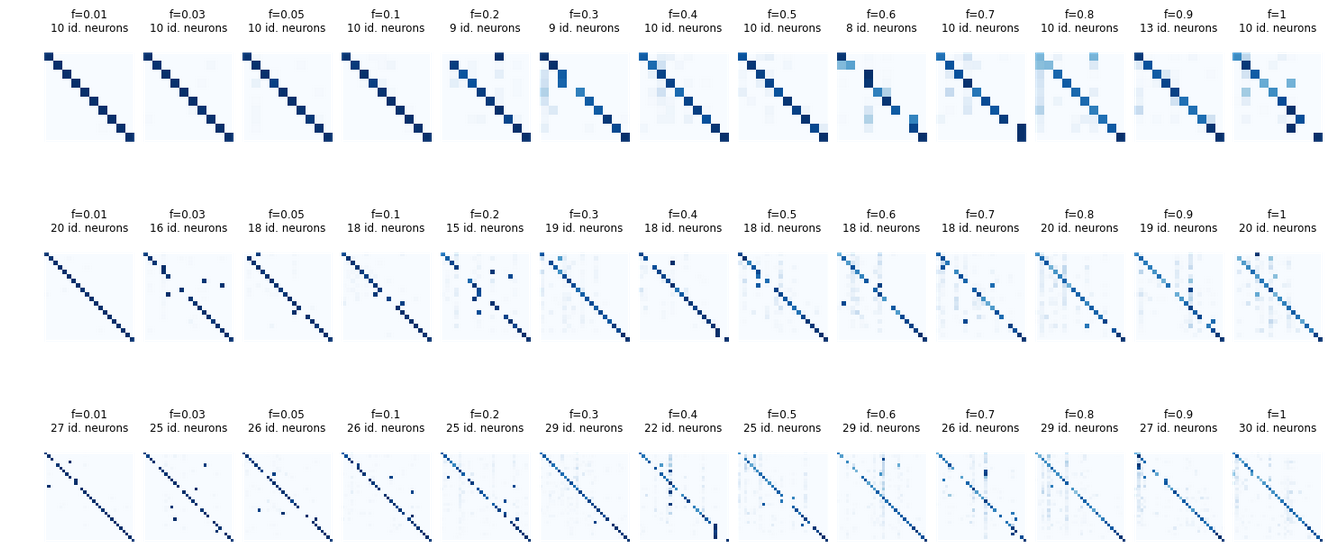}
\caption{Simulation results for amplitudes between $0$ and $20$. The number of neurons vary: on the first row we take $10$ neurons, on the second row we take $20$ neurons and on the third row we take $30$ neurons. On each row, the superposition frequence $f$ increases from left to right: $0.01$, $0.03$, $0.05$, $0.1$, $0.2$, $0.3$, $0.4$, $0.5$, $0.6$, $0.7$, $0.8$, $0.9$, $1$. ``Identified neurons" is abbreviated by ``id. neurons".}
\label{Change_freq20}
\end{figure}

In Figure \ref{Change_freq20}, we show the results for amplitudes between $0$ and $20$. We take $10$, $20$ and $30$ neurons, and for each fixed number of neurons we vary the superposition frequence. We can see that the confusion matrices show a lot of weight on the diagonal, even with a high superperposition frequence. We can conclude that in this setting with high amplitudes, ToMATo is extremely robust to superpositions. \\

\begin{figure} 
\centering      
\includegraphics[scale=0.3]{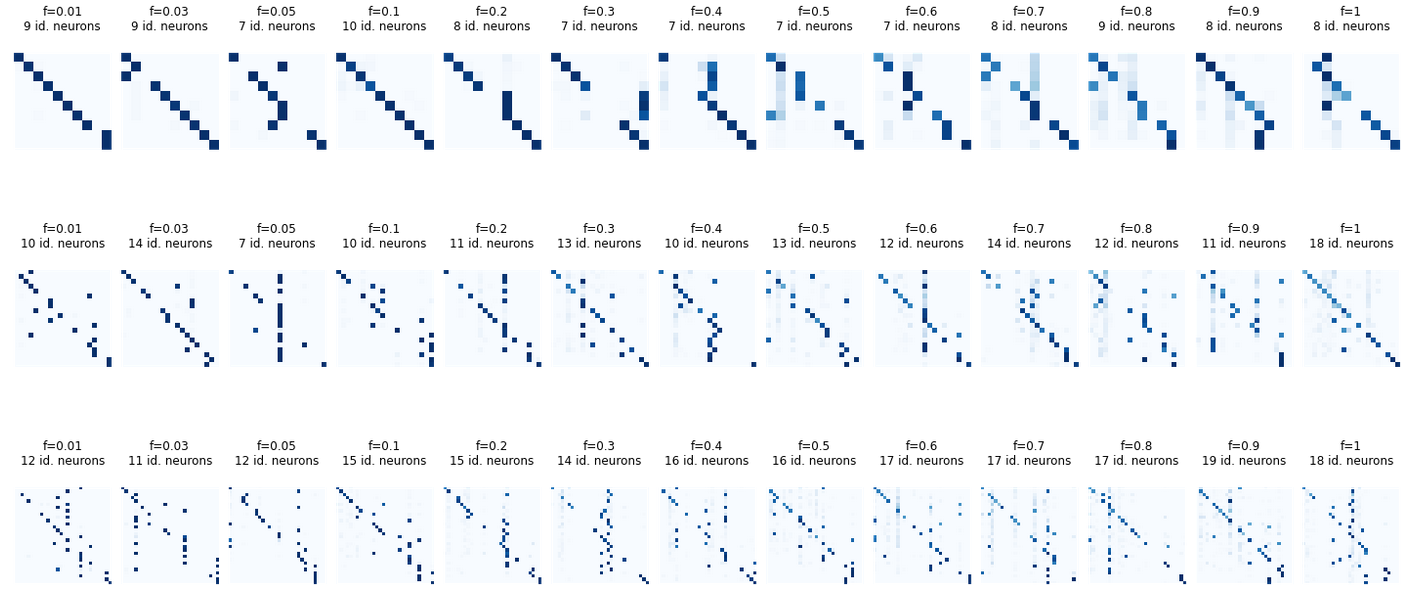}
\caption{Simulation results for amplitudes between $0$ and $10$. The number of neurons vary: on the first row we take $10$ neurons, on the second row we take $20$ neurons and on the third row we take $30$ neurons. On each row, the superposition frequence increases from left to right: $0.01$, $0.03$, $0.05$, $0.1$, $0.2$, $0.3$, $0.4$, $0.5$, $0.6$, $0.7$, $0.8$, $0.9$, $1$. ``Identified neurons" is abbreviated by ``id. neurons".}
\label{Change_freq10}
\end{figure}

Simulations that give Figure \ref{Change_freq10} are performed in the same setting but for amplitudes between $0$ and $10$. Results are a little bit less satisfying but once again, there is no clustering method that could perfectly detect neurons with events of low amplitude. Moreover, we can observe is this case too, that the results stay consistent when the superposition frequency increases.\\

From these simulations, we can conclude that: 
\begin{itemize}
\item As any clustering method, ToMATo is not very robust to low amplitudes, but it still gives very satisfying results.
\item ToMATo is extremely robust to the presence of superpositions, which is very impressive. 
\end{itemize}

\subsection{Application on real data}\label{appli}
We now apply ToMATo on a real locust dataset.

\subsubsection{Description of the data}
The data used here were recorded from the ﬁrst olfactory relay, the antennal lobe, of
a locust (\emph{Schistocerca americana}). Recording setting and acquisition details are described in \cite{Using_noise}. Measurements are performed using a tetrode, thus neuronal activity is recorded at $4$ different sites. \\
 
\begin{figure} 
\centering
\includegraphics[scale=0.3]{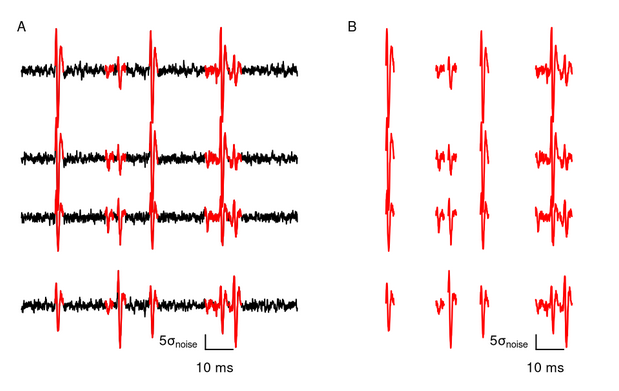}
\caption{Recordings data and event detection.}
\label{tri}
\end{figure}

In Figure \ref{tri} A, we show a 100 ms of data recorded at the 4 sites of a tetrode. The data were filtered between 300 Hz and 5 kHz before being digitized with a sampling frequency of 15 kHz. They were then normalized by dividing the signal amplitude at each site by a robust estimator, the median absolute deviation, of the noise deviation $\sigma_{\mathrm{noise}}$. In Figure \ref{tri} B, after detection of the spike candidates as sufficiently large local extrema in absolute value, cuts are made on each of the four sites, each cut having $45$ sampling points. This group of four cuts determines an event, of dimension $4 \times 45=180$. The procedure is detailed in \cite{SpySort} and summarized in Appendix \ref{annexe}.\\

\subsubsection{Result of ToMATo}

\begin{figure}  
\centering 
\includegraphics[scale=0.2]{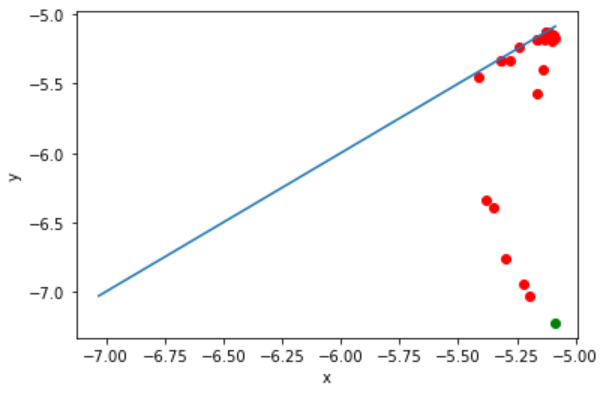}
\includegraphics[scale=0.3]{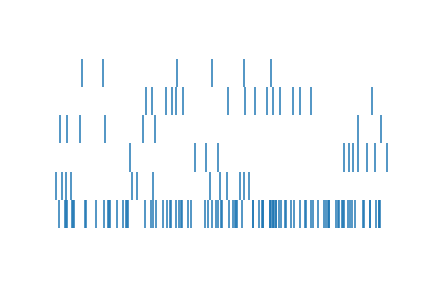}
\caption{Left: Persistence diagram. Right: The $2$ first seconds of the final obtained rasterplot.}
\label{dgm_raster}
\end{figure}

Let us apply ToMATo to this event collection. The first run of ToMATo gives the persistence diagram displayed in Figure \ref{dgm_raster} Left. There are clearly $6$ points far away from the diagonal thus we run ToMATo again with n\_clusters$=6$. In Figure \ref{dgm_raster} Right, we display the obtained rasterplot. Each horizontal row corresponds to one of the $6$ identified neurons, and shows a series of the times at which this neuron emitted a spike.

\begin{remark}
With the usual method (presented in \url{https://c_pouzat.gitlab.io/spike-sorting-the-diy-way/} and summarized in Appendix A), with the same dataset it is not clear whether one should choose $5$ or $6$ clusters. With the ToMATo method there is no ambiguity. 
\end{remark} 
 
The code and the data used in this article are available at \url{https://gitlab.math.unistra.fr/martineau/tomato_for_spike_sorting}. 
  
\section{Conclusion} 

The ToMATo clustering algorithm (Sec.~\ref{sec:tomato}) is part of a recent branch of applied mathematics, topological data analysis (TDA) \cite{Topo_and_data, Simplification}; a branch that grew out of a rather sophisticated and abstract domain of mathematics, algebraic topology. Despite of this abstract origin, the key ideas on which ToMATo is built can (and should, we think) be grasped by any serious practitioner of neurophysiological data analysis. The introduction and use of the concept of \textbf{peak prominence} (Sec.~\ref{sec:persistent-homology}) gives rise to a mode-seeking algorithm that is \textbf{robust with respect to noise}. The use of a graph built from the estimated density values at the data points, \textbf{and only at the data points}, leads to an  algorithm that is computationally very efficient (Sec.~\ref{sec:tomato-explanation}). Serious mathematical studies have moreover provided theoretical guarantees (Sec.~\ref{sec:tomato-principle}) on the capabilities of ToMATo to recover the right clusters. These considerations convinced us that an exploration of ToMATo in a neurophysiological data analysis context was a worthy endeavor.\\

As a test case, we chose a subject we know reasonably well: spike-sorting (Sec.~\ref{app}). More precisely, the part of spike sorting where clustering algorithms play a key role is the determination of the number of ``good'' neurons---a good neuron is a neuron whose spikes can be reliably identified---, together with their waveform/template (Sec.~\ref{sec:spike-sorting-short} and Appendix~\ref{annexe}). For our test, both simulated (Sec.~\ref{sim_res}) and real data (Sec.~\ref{appli}) were used. The very important preliminary question, the determination of the number of clusters/neurons, was discussed twice; first in the general presentation of ToMATo (Sec.~\ref{expl} and \ref{tau}), next in the specific context of spike sorting (Sec.~\ref{sec:number-of-neurons}). We considered two broad simulation scenarios : ``locust'', the easy case, with a large signal to noise ratio (SNR), few neurons, leading to few superpositions; and ``hippocampus'', with a low SNR and many neurons, leading to frequent superpositions. We showed that ToMATo does not require a preliminary dimension reduction and is able to reliably identify the ``good'' neurons \textbf{despite of a potentially large number of superposed events}. These findings were confirmed using our ``usual'' real dataset.\\ 

We hope that this report has convinced our readers that the ToMATo clustering algorithm provides an attractive alternative to more traditional algorithms (kmeans, Gaussian mixture models, etc.). ToMATo is moreover just a tiny part of a mature, comprehensive and well documented topological data analysis \texttt{C++} library: Gudhi. The library is interfaced with \texttt{Python} (and \texttt{R}) and can be readily tried by anyone willing to spend a few time reading the relevant documentation. Don't hesitate!
 
\section*{Acknowledgments} 
The authors would like to thank Marc Glisse from INRIA for carefully answering our questions about the ToMATo implementation. 
This work was supported by the Agence Nationale de la Recherche (ANR): project ANR-22-CE45-0027 SIMBADNESTICOST.
 
\appendix

\section{A spike sorting outline}\label{annexe} 

\begin{figure} 
  \centering
  \includegraphics[width=0.9\textwidth]{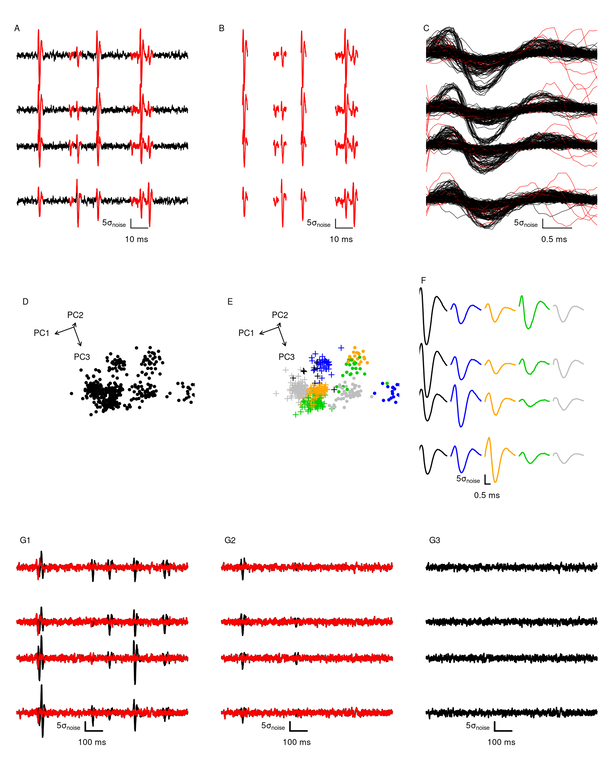}
  \caption{\label{sorting-outiline}Spike sorting as a succession of ``simple'' tasks (see text and \url{https://c_pouzat.gitlab.io/spike-sorting-the-diy-way/} for details). These images are from \url{https://c_pouzat.gitlab.io/spike-sorting-the-diy-way/}.}
\end{figure}

Spike sorting aims at extracting from ``raw data'' (nowadays continuous recordings from several channels/electrodes, like the 4 channels of Fig.~\ref{sorting-outiline}A) sequences of spike times emitted by ``identified'' neurons \cite{einevoll.ea:12,pouzat:16,leblois.pouzat:17}. The raw data are typically a mixture of waveforms/motifs from different neurons with an added independent recording noise. The data generation model that is most of the time tacitly assumed, but sometimes spelled out \cite{roberts:79}, is that the neurons generate dependent marked point processes \cite{bremaud:20}, where the mark of each neuron on each channel/electrode is a waveform/motif; when two or more neurons generate events whose time separation is smaller than the duration of their waveforms, these waveforms are summed on their overlapping regions. A white or ``colored'' noise is then added to the realizations of these marked point processes. We wrote that the raw data are typically \textbf{continuous} recordings, but this descriptive statement should not mislead the mathematically oriented reader: the data are in fact \textbf{sampled}---that is, measurements are performed at a fixed frequency (15 kHz for the data of Fig.~\ref{sorting-outiline}), implying that the data are intrinsically discrete, they just ``look'' continuous---and they are filtered before being sampled, a high-pass filter (300 Hz for the data of Fig.~\ref{sorting-outiline}) removes low frequency oscillations and a low-pass filter (5 kHz for the data of Fig.~\ref{sorting-outiline}) ensuring that the requirement of Nyquist-Shannon sampling theorem \cite{unser:00} are met (the sampling frequency must be at least twice as large as the low-pass filter cut-off frequency). Most of the (really) many spike sorting algorithms that have been proposed can be decomposed in 3 main stages illustrated by the 3 rows on Fig.~\ref{sorting-outiline}:
\begin{enumerate}
\item Spike detection (Fig.~\ref{sorting-outiline}A, in red), and event sample construction (Fig.~\ref{sorting-outiline}B and C).
\item Estimation of the number of neurons and of their waveforms (Fig.~\ref{sorting-outiline}D, E and F).
\item Event assignment and superposition resolution (Fig.~\ref{sorting-outiline}$\mathrm{G}_1,\mathrm{G}_2,\mathrm{G}_3$).
\end{enumerate}
The present manuscript focuses on the second point. For completeness we state below the main operations we usually apply to the data in order to provide a clear ``reference'' procedure with which our new approach can be compared (referring to Fig.~\ref{sorting-outiline})\footnote{The data and \texttt{Python} codes leading to this figure are publicly available: \url{https://c_pouzat.gitlab.io/spike-sorting-the-diy-way/}.}:
\begin{itemize}
\item[A] Events detection (putative action potentials) based on extrema exceeding a threshold.
\item[B] Cuts / windows, one on each site, of ``well-chosen'' length (here 45 sampling points) around the detected extremes, this collection of four cuts (as we have here four sites) constitutes \textbf{an event} (our event space is here $\mathbb{R}^{180}$ as we have 4×45 amplitudes per event).
\item[C] The first 200 detected events aligned on their valley (events made of superpositions are displayed in red).
\item[D] \textbf{Dimension reduction}, here the projection of the sample on a plane of the subspace defined by the first three principal components.
\item[E] \textbf{Clustering} with the k-means method and 10 centers.
\item[F] The motifs (centers of the clusters defined in the previous step) corresponding to the 5 ``largest'' neurons (the 5 different colors) on each of the 4 sites.
\item[G1] Return to the raw data (black trace) and attribution of a motif to each local extremum generating a prediction (red trace).
\item[G2] In black the difference between the black and red traces of G1, a detection of local extrema is performed again and the closest motif is assigned to each extremum, giving rise to a new prediction (red trace).
\item[G3] The difference between the black and red traces of G2, we continue this ``peeling'' procedure until there is nothing left identifiable to any of the motifs of the collection.
\end{itemize}
This somewhat long list should be taken as \textbf{an example} of what is done when doing spike sorting, since there are usually several options/approaches at each step. We use for instance PCA for dimension reduction \cite{glaser.marks:68}, but ICA can be (and is) used. The ``peeling procedure'' of the third row \cite{pillow.ea:13,SpySort}, a sophisticated form of \textbf{template matching}, is just one way of solving the superposition problem \cite{prochazka.conrad.sindermann:72}. The fact that the data are sampled gives rise to a \textbf{sampling jitter} \cite{mcgill.dorfman:84} that must be dealt with in order to get good superposition resolution. This sampling jitter together with the presence of superpositions makes spike sorting a \textbf{non trivial clustering problem}. As explained in the Introduction, the superpositions must be eliminated in an ad-hoc way in order to have useful principal components when dimension reduction is used. We are not dealing here with the added difficulties met in some datasets where the waveform of a given neuron depends on the history of its discharge \cite{calvin:73}; using closely spaced recording electrodes and working with the amplitude ratios on the different site is usually the best way to proceed \cite{mcnaughton.okeefe.barnes:83,gray.ea:95}. 

\newpage
\bibliographystyle{plain}
\bibliography{biblio}
 
\end{document}